\documentclass[]{spie}  

\usepackage{amsmath,amsfonts,amssymb}
\usepackage{graphicx}
\usepackage[colorlinks=true, allcolors=blue]{hyperref}
\usepackage{makecell}
\usepackage{colortbl}
\usepackage{array}
\usepackage{multirow}

\usepackage[table]{xcolor}

\title{Understanding the effects of charge diffusion in next-generation soft X-ray imagers}

\author[a]{Eric~D.~Miller}
\author[a]{Gregory~Y.~Prigozhin}
\author[a]{Beverly~J.~LaMarr}
\author[a]{Marshall~W.~Bautz}
\author[a]{Richard~F.~Foster}
\author[a]{Catherine~E.~Grant}
\author[b]{Craig~S.~Lage}
\author[c]{Christopher~Leitz}
\author[a]{Andrew~Malonis}

\affil[a]{Kavli Institute for Astrophysics and Space Research, Massachusetts Institute of Technology, 77 Massachusetts Ave, Cambridge, MA 02139, USA}
\affil[b]{Department of Physics, UC Davis, One Shields Avenue, Davis, CA 95616, USA}
\affil[c]{Lincoln Laboratory, Massachusetts Institute of Technology, 244 Wood St., Lexington, MA 02421, USA}

\authorinfo{Further author information:\\E.~D.~Miller: E-mail: milleric@mit.edu}

\begin{document} 
\maketitle

\begin{abstract}
To take advantage of high-resolution optics sensitive to a broad energy range, future X-ray imaging instruments will require thick detectors with small pixels. This pixel aspect ratio affects spectral response in the soft X-ray band, vital for many science goals, as charge produced by the photon interaction near the entrance window diffuses across multiple pixels by the time it is collected, and is potentially lost below the imposed noise threshold. In an effort to understand these subtle but significant effects and inform the design and requirements of future detectors, we present simulations of charge diffusion using a variety of detector characteristics and operational settings, assessing spectral response at a range of X-ray energies. We validate the simulations by comparing the performance to that of real CCD detectors tested in the lab and deployed in space, spanning a range of thickness, pixel size, and other characteristics.  The simulations show that while larger pixels, higher bias voltage, and optimal backside passivation improve performance, reducing the readout noise has a dominant effect in all cases. We finally show how high-pixel-aspect-ratio devices present challenges for measuring the backside passivation performance due to the magnitude of other processes that degrade spectral response, and present a method for utilizing the simulations to qualitatively assess this performance. Since compelling science requirements often compete technically with each other (high spatial resolution, soft X-ray response, hard X-ray response), these results can be used to find the proper balance for a future high-spatial-resolution X-ray instrument.

\end{abstract}

\keywords{X-ray detectors, CCDs, charge diffusion, detector response}

\section{Introduction}
\label{sect:intro}

\begin{figure}[t]
\begin{center}
\includegraphics[width=.8\linewidth]{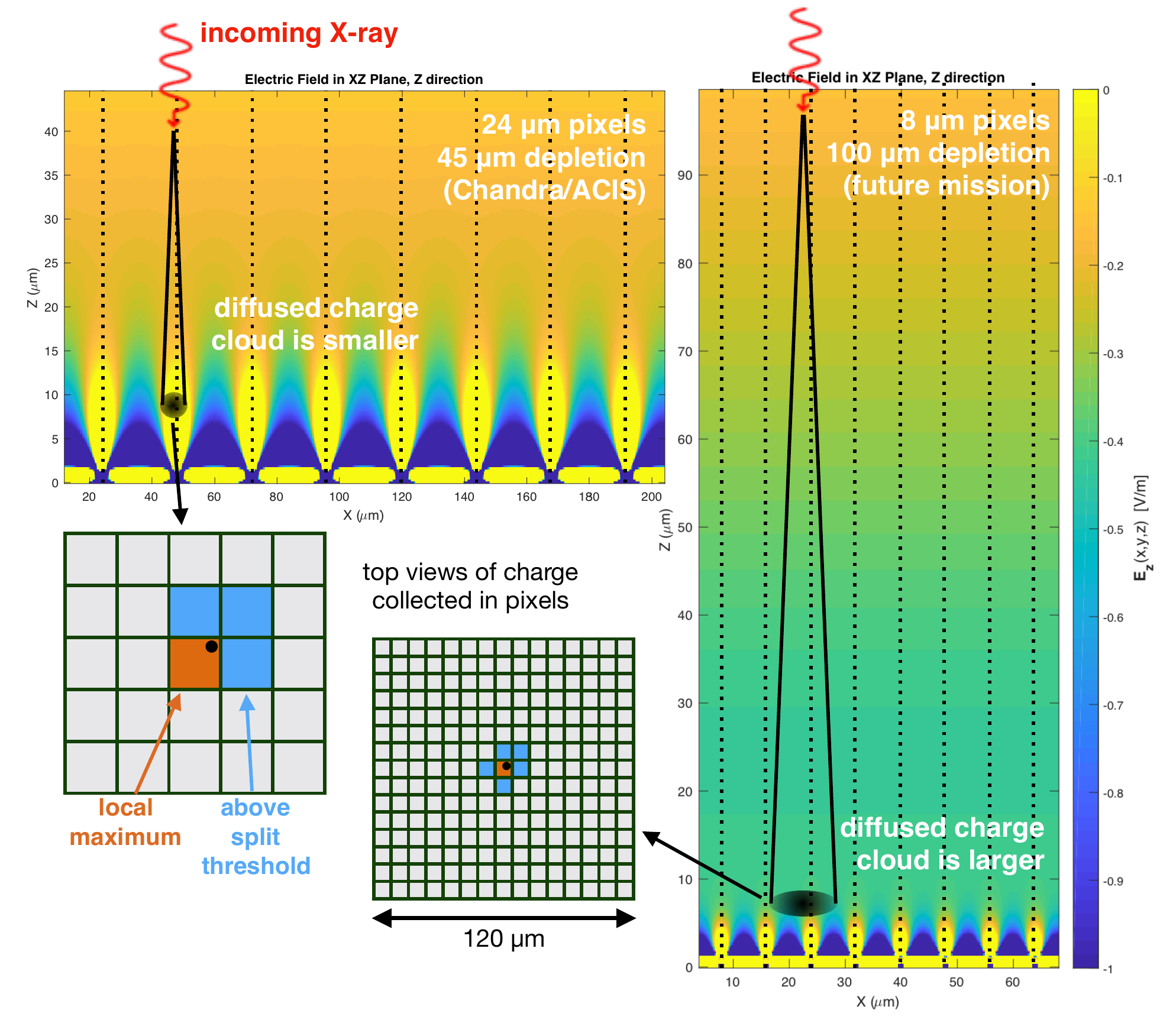}
\end{center}
\caption{The effects of thicker detectors with smaller pixels on X-ray event energy reconstruction. The colored panels show the vertical electric field strength in two simulated back-illuminated silicon detectors: a 45-$\mu$m thick device with 24-$\mu$m pixels, similar to Chandra ACIS-S3 or Suzaku XIS1; and a 100-$\mu$m thick device with 8-$\mu$m pixels, similar to what might be flown on a future mission. A soft X-ray interacting near the entrance window has a longer distance to drift in the thicker detector, resulting in more charge diffusion over a larger number of smaller pixels, as shown in the top-down view of pixel grids in the lower left.}
\label{fig:fields}
\end{figure}

Future silicon-based soft X-ray imaging instruments will require small pixels to take advantage of advanced high-resolution optics, while also greatly benefiting from thick detector bulk to maximize the sensitive band-pass. For example, the wide-field imaging detectors baselined for both the AXIS Probe-class mission concept\cite{AXIS} and the Lynx Great Observatory concept\cite{Gaskinetal2017} require $\sim$16-$\mu$m pixels to properly sample the $1^{\prime\prime}$ PSF goal, yet they need to be  backside-illuminated devices fully depleted to $\sim$100 $\mu$m to provide sufficient quantum efficiency over the full 0.2--12 keV band. This pixel aspect ratio of $\sim$6:1 is unprecedented among similar X-ray instruments, which are closer to 2:1 for Chandra ACIS-S3 and Suzaku XIS1. ``Taller'' pixels allow for more lateral diffusion of the X-ray-induced photoelectrons as they drift toward the collection gates under influence of the bias electric field, as demonstrated schematically in Figure \ref{fig:fields}. This diffusion is greatest for soft photons which interact with the silicon bulk near the entrance window, and since these photons also produce fewer photoelectrons than their harder counterparts, charge can be spread out across many pixels and potentially lost below the imposed noise thresholds, having substantial effects on the overall soft X-ray response. Here ``response'' refers to the combined effects of our ability (or lack thereof) to reconstruct the energy of individual detected photons, and mainly includes the spectral resolution FWHM, knowledge of the energy scale, and reduced quantum efficiency due to events lost below the noise threshold. 

It is at the softest energies where some of the most ground-breaking high-energy astrophysics will be done, including probing the seeds of the first supermassive black holes at high redshift, tracing the faintest hot gas in the outskirts of galaxy clusters and the tenuous intergalactic medium, and understanding the role of stellar and AGN feedback in shaping galaxies like our own Milky Way\cite{AXIS,Gaskinetal2017}\footnote{See the Concept Study Reports for Lynx (\url{https://wwwastro.msfc.nasa.gov/lynx/docs/LynxConceptStudy.pdf}) and AXIS (\url{https://smd-prod.s3.amazonaws.com/science-red/s3fs-public/atoms/files/AXIS\_Study\_Rpt.pdf}).}. For these reasons, understanding the processes governing the movement of charge in silicon detectors is an area of active research.\cite{Prigozhinetal2003,Miller2018,Haroetal2020,Prigozhinetal2021,LaMarretal2022,Avalosetal2022} In this work, we extend our recent investigation of charge diffusion\cite{LaMarretal2022} in real detectors under development for strategic missions like AXIS and Lynx\cite{Bautzetal2022} by using simulations. These are validated with real data from different sources, and we use the results to explore the effects of different detector design parameters on the X-ray response across a range of energies from 0.3 to 6 keV. We emphasize that these results are applicable to any silicon-based X-ray imager, including CCDs\cite{Bautzetal2022}, CMOS devices\cite{Chattopadhyayetal2018}, and DEPFETs\cite{Meidinger2017}.

\section{Simulations and data analysis}
\label{sect:data}

\subsection{Simulating the detector electric field and charge diffusion}
\label{sect:simulations}

\begin{figure}[t]
\begin{center}
\includegraphics[width=.75\linewidth]{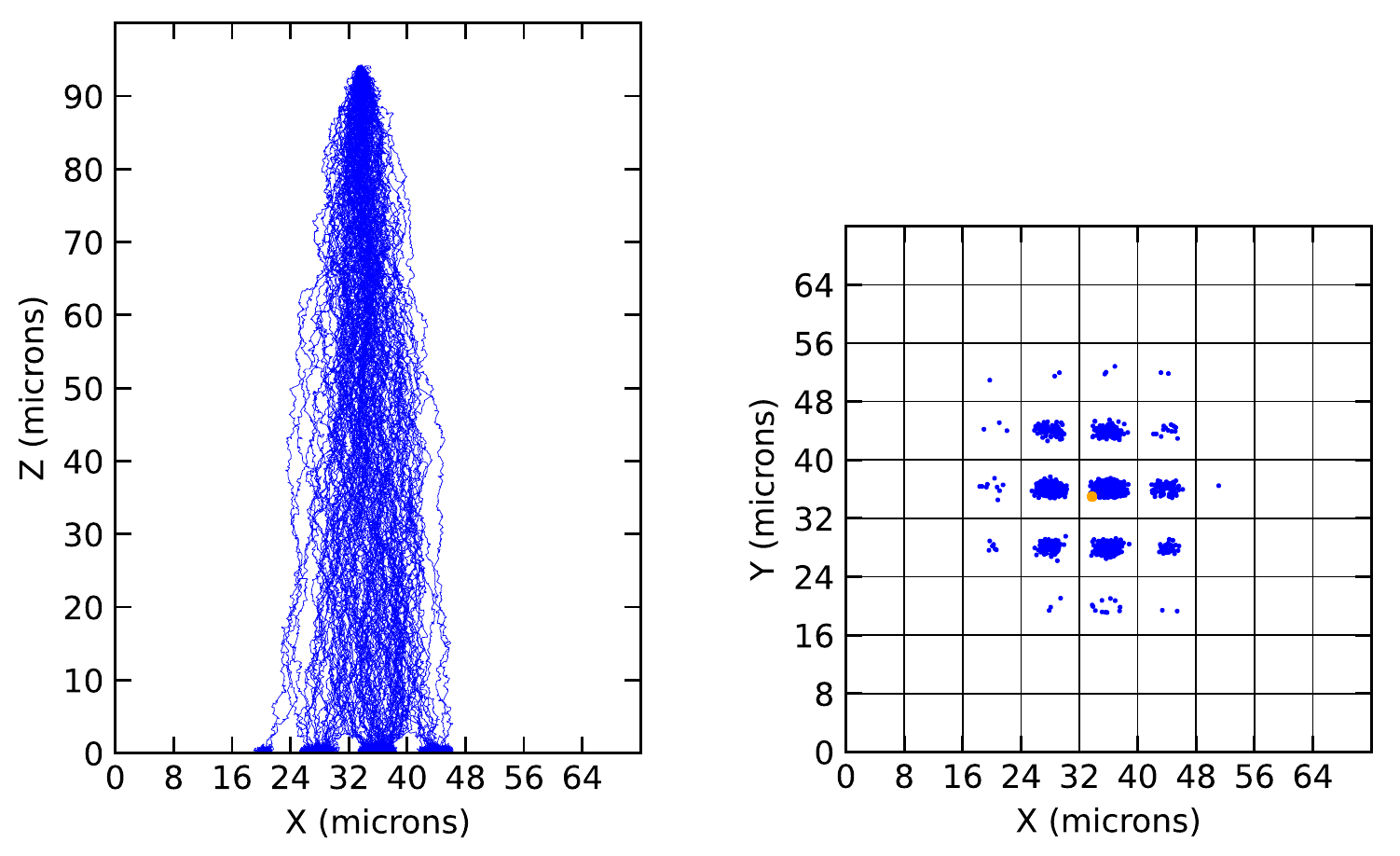}
\end{center}
\caption{An example run tracking the charge diffusion in \textsc{Poisson CCD} for a single 5.9 keV Mn K$\alpha$ photon interacting $\sim$6 $\mu$m from the entrance window in a device with $V_\mathrm{sub} = -50$ V. The left panel shows tracks for individual photoelectrons as they drift and diffuse in the device's electric field. For clarity, only 100 of the $\sim$1600 electron tracks are shown. Note the separation into 8-$\mu$m pixels near $Z=0$. The right panel shows a top view of the final electron locations, with all $\sim$1600 electrons represented by blue dots. The orange dot indicates the projected $X,Y$ location of the initial photon interaction. With 8-$\mu$m pixels, signal ends up in many pixels, with several containing only a handful of electrons.}
\label{fig:tracks}
\end{figure} 

We performed simulations using \textsc{Poisson CCD} \cite{Lageetal2021}, a software package that, given a set of structural and operational characteristics for the detector, (1) solves Poisson's equation for the electric field in three dimensions at all points within the detector volume, and (2) tracks the drift and diffusion of electrons placed within this volume until they reach the buried channel, where they are collected in pixels defined by the channel stop and barrier gate fields. The simulated detector is a backside-illuminated, 100-$\mu$m thick CCD with 8-$\mu$m pixels, a three-phase gate structure with one collecting gate held at +1.5 V and two barrier gates held at $-$1.5 V, and implant structure and doping levels based on the MIT Lincoln Lab CCID-93 device described in our recent comparison of lab data and simulation\cite{LaMarretal2022}. The simulation volume covered 9$\times$9 pixels, with non-linear electric field grid spacing in the vertical direction allowing finer grid sampling of 30 nm at the front and back sides of the device to properly capture the electric field structure, and coarser grid sampling (up to 300 nm) throughout the bulk of the device. The electric field was simulated at an operating temperature of $-50 ^{\circ}$C at substrate bias voltages of $V_\mathrm{sub} = -30$, $-50$, and $-100$ V, all sufficient to fully deplete the silicon bulk and eliminate field-free regions. Examples of vertical electric field structure resulting such simulations are shown in Figure \ref{fig:fields}.

Charge diffusion was simulated for each $V_\mathrm{sub}$ using 100,000 photons each at energies of 5.89 keV (Mn K), 1.25 keV (Mg K), 0.525 keV (O K), and 0.277 keV (C K). This was done by introducing small clouds of electrons into the simulated detector volume and allowing them to drift and diffuse under the influence of the 3-D electric field until collected in pixels. The number of electrons in each cloud was drawn from a Fano noise distribution appropriate for the photon energy in Si (e.g., 1615$\pm$14 electrons for 5.9 keV), and the interaction depth was drawn from an exponential distribution with appropriate energy-dependent attenuation length. Each interaction location was generated from a uniform distribution across the central pixel of the 9$\times$9-pixel simulation volume. An example diffusion run is shown in Figure \ref{fig:tracks}. 

We simulated the quality of backside passivation by adjusting the \textsc{TopAbsorptionProb} parameter, which is the probability that a diffused photoelectron encountering the backside entrance window is absorbed rather than reflected back into the bulk; this ranged from 0\% (perfect passivation) to 25\%. The final pixelized 2-D spatial distribution of electrons was binned 2$\times$2 and 3$\times$3 to simulate 16-$\mu$m and 24-$\mu$m pixels, respectively; although this is technically different than simulating drift and diffusion in a detector with the correct pixel size, it greatly reduced the required simulation time, and several test runs showed there was no noticeable difference between the methods, as the majority of the diffusion occurs in the bulk before the charge cloud feels the effects of the gate barrier and channel stop electric fields (see Figures \ref{fig:fields} and \ref{fig:tracks}, left panel). Finally, Gaussian readout noise ranging  from 1 to 4 e$^-$ RMS was added to each pixel, and event detection and characterization were performed in the same way as flight software on Chandra ACIS and Suzaku XIS, as described in our previous work.\cite{LaMarretal2022} For each single-photon frame, we used a 5\,$\sigma$ threshold for event candidate identification, and summed all pixels including neighbors above 4\,$\sigma$ to calculate the total event energy. These thresholds are similar to values tuned for previous instruments to eliminate noise events and inclusion of noisy pixels in event pulse height summation. At a temperature of $-50^{\circ}$C, we expect a conversion of 3.71 eV per e$^-$\cite{Groometal2006}, and so in the highest noise case of 4 e$^-$ RMS, the event threshold corresponds to 74 eV and the neighbor (or ``split'') threshold to 59 eV. These are high enough to lose a significant fraction of the signal in a low-energy event split over several pixels. In addition, a 5\,$\sigma$ event threshold has a 25\% chance of introducing a noise event in a 1-Mpix array; for the expected $\sim$20 fps frame rate for a mission like AXIS, this will introduce many such events, and thus an event threshold that is many factors of the readout noise and still well below the science band of interest is desirable.

As also discussed in our previous work\cite{LaMarretal2022}, the default charge diffusion parameters in \textsc{Poisson CCD} produced much more diffusion in the simulations than we observe in the lab data, with much larger event sizes. We used similar \textsc{DiffMultiplier} parameter values that produced simulated event size distributions closely matched to the real data, and also similar to values supported by theoretical studies. The validation with real data presented in our previous work and here in Section \ref{sect:validation} support use of this diffusion factor, and in fact more diffusion would magnify the importance of factors such as readout noise, pixel size, and bias voltage in the resultant soft response. The slight discrepancy between our tuned \textsc{DiffMultiplier} value and the expected value will be explored in a future paper.

\begin{figure}[t]
\begin{center}
\includegraphics[width=.7\linewidth]{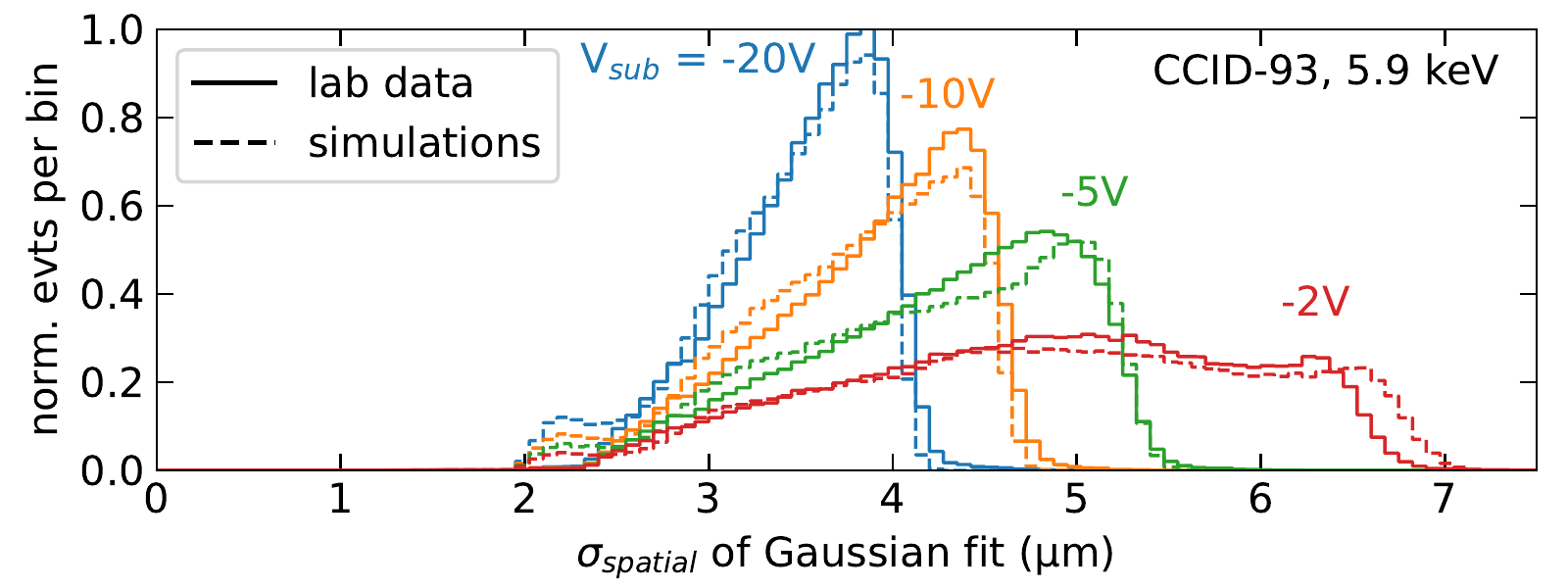}
\includegraphics[width=.7\linewidth]{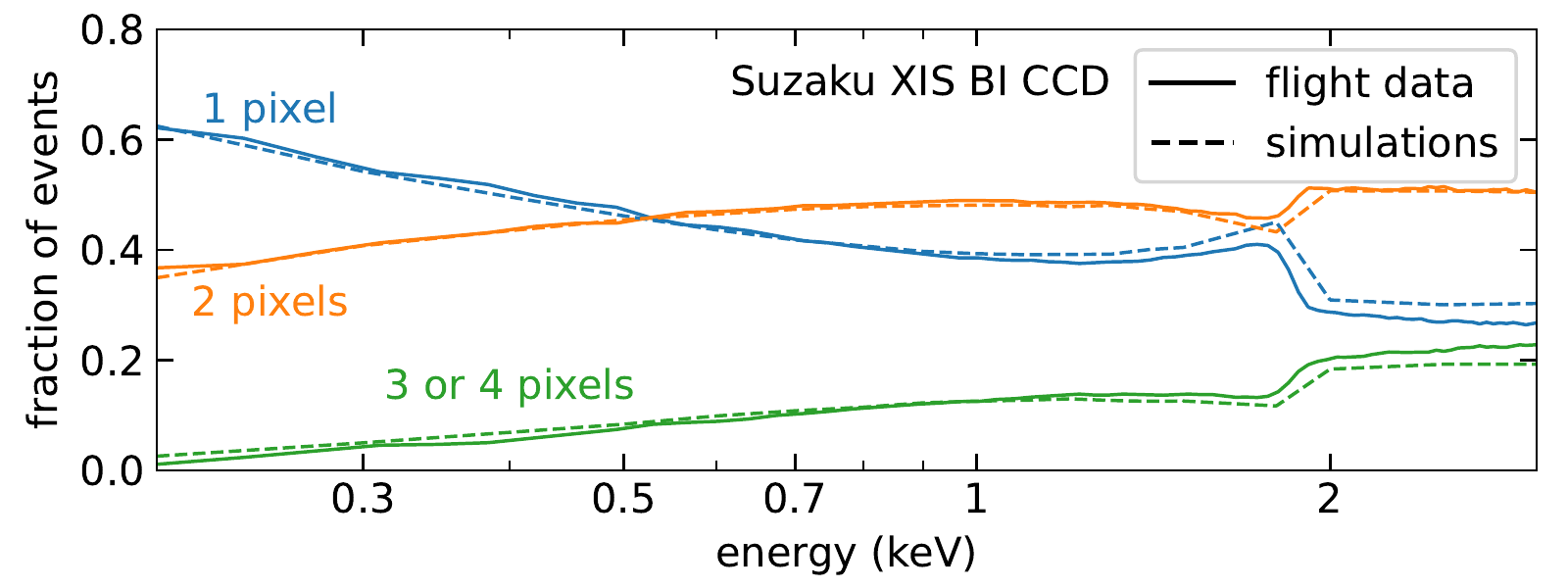}
\end{center}
\caption{Validation of the charge diffusion simulations with real data. The top panel compares lab data using a very similar but thinner detector to that used in our other simulations here. As described in our previous work\cite{LaMarretal2022}, we fit a 2-D Gaussian to the pixel islands of the measured and simulated events at a range of substrate bias voltage $V_{\mathrm sub}$, and find that the distribution of Gaussian $\sigma$ (a proxy for lateral diffusion magnitude) is similar. The bottom panel shows the fraction of events spanning 1, 2, or 3--4 pixels in real Suzaku XIS1 flight data compared to simulations of that same device. The fractions match remarkably well at these soft energies, suggesting that the charge diffusion is accurately simulated.}
\label{fig:validation}
\end{figure} 

\subsection{Validation of simulations with real data}
\label{sect:validation}

To validate the simulations, we simulated charge diffusion in two real X-ray detectors for which we have substantial data, both BI CCDs fabricated by MIT Lincoln Laboratory: the CCID-93 small-format, small-pixel (8-$\mu$m) device\cite{LaMarretal2022}; and the CCID-41 device flown as Suzaku XIS1\cite{Koyamaetal2007}.  The CCID-93 simulations are similar to the unbinned 5.9-keV runs described in Section \ref{sect:simulations}, although with 50-$\mu$m thickness, and as in our previous work\cite{LaMarretal2022}, we fit 2-D Gaussians to pixelated event islands to measure the charge diffusion. As shown in the top panel of Figure \ref{fig:validation}, the simulated sizes match the lab data very well. The Suzaku data includes flight data from a number of bright, un-piled sources, including the Perseus Cluster and the bright blazars PKS 2155 and 3C 273, the latter taken in 1/8 window mode. This data from these bright continuum sources is dominated by X-ray photons across a wide energy band, with very little contamination from particle events. We parametrize charge diffusion using the ASCA grade specifying how many pixels in an event are above the noise threshold. As shown in the bottom panel of Figure \ref{fig:validation}, the simulations match the observed charge diffusion exceptionally well at soft energies. While both real devices are $\sim$50 $\mu$m thick, about half that of the notional simulated detector, the dependence of diffusion with drift distance is well-defined, and this validation gives us confidence that our simulation results are reliable.

\section{Results}
\label{sect:results}

\subsection{Spectral response}
\label{sect:response}

\begin{figure}[p]
\begin{center}
\includegraphics[width=.70\linewidth]{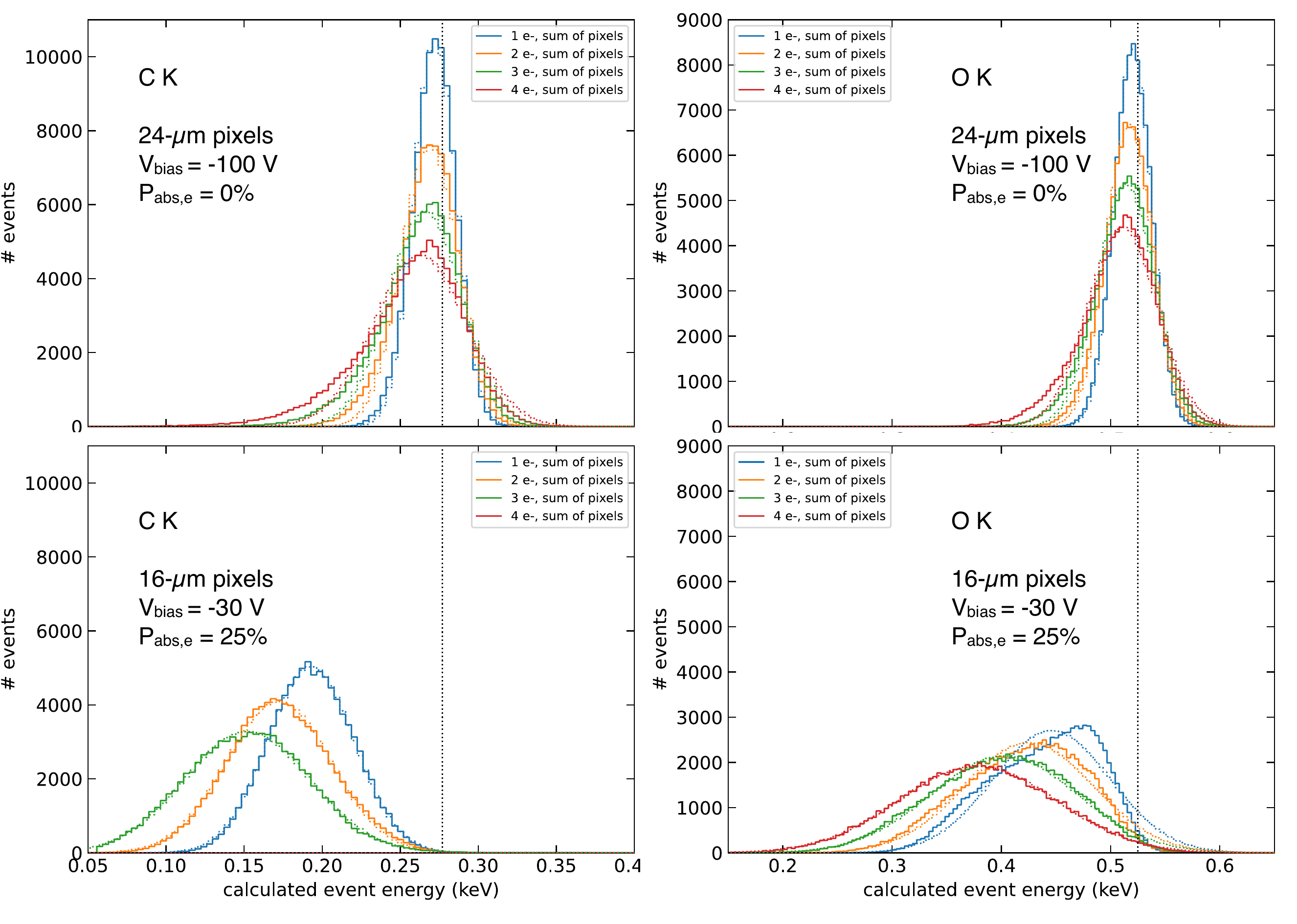}
\end{center}
\caption{Spectral response of the simulated detector to monochromatic photons of (left) C K at 0.28 keV and (right) O K at 0.53 keV. Top panels show high-quality charge collection with large pixels, high bias voltage, and perfect backside passivation. Bottom panels show less-than-ideal charge collection, with small pixels, low bias voltage, and imperfect passivation, diminished performance that may be imposed by other instrument constraints. The event energies are calculated by summing pixels above threshold, as described in the text. Higher readout noise increases the spectral FWHM in all cases, and in the less-than-ideal case causes the histogram to shift to much lower energies and all events to be lost in the worst case at C K.}
\label{fig:spec}
\end{figure}

\begin{figure}[p]
\begin{center}
\includegraphics[width=.70\linewidth]{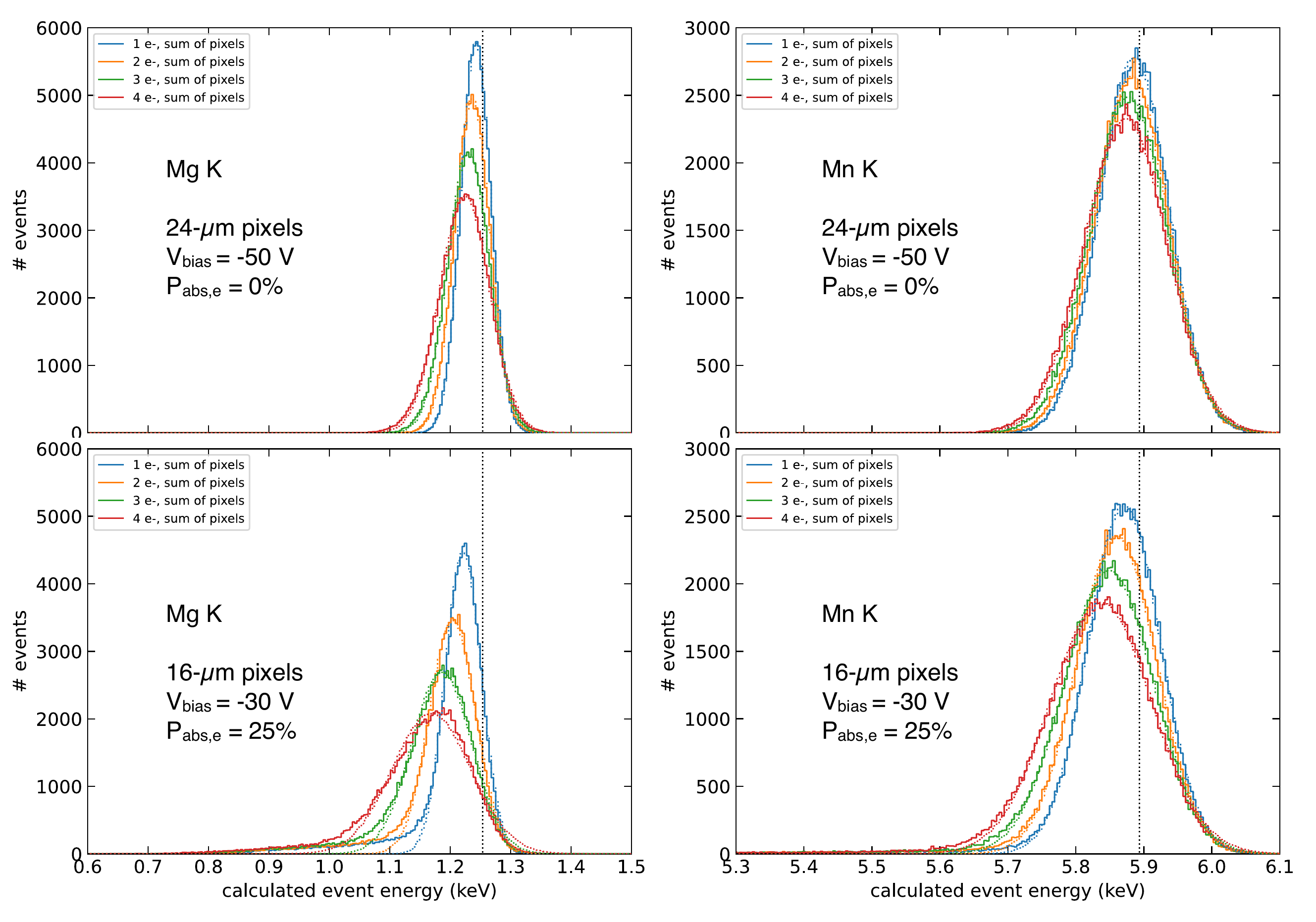}
\end{center}
\caption{Same as Figure \ref{fig:spec}, but for (left) Mg K at 1.25 keV and (right) Mn K at 5.9 keV. The effects of the less-than-ideal detector and higher readout noise are moderate at these energies, although non-Gaussian features are obvious in the higher noise spectra.}
\label{fig:spec_mn_mg}
\end{figure}

The effects of charge diffusion on spectral response are apparent in all of our simulations. Figures \ref{fig:spec} and \ref{fig:spec_mn_mg} show histograms of reconstructed event energy for the four monochromatic energies under study: C K (0.28 keV), O K (0.53 keV), Mg K (1.25 keV), and Mn K (5.9 keV). Colors show results for different values of readout noise, ranging from 1 to 4 e$^-$ RMS. In each figure, an ``ideal'' detector from the standpoint of maximizing charge collection is shown in the top panels. A lower performance but still realistic detector is represented in the lower panels. Pixel size, substrate bias voltage, and back surface passivation quality contribute to the width and shape of the response. These effects are greatest at low energies, where peak energy shifts and non-Gaussian shapes appear even for low readout noise. Quantum efficiency (QE) is affected as well, in the worst case resulting in no events being detected for C K with 4 e$^-$ RMS noise (Figure \ref{fig:spec}, lower left).

\begin{figure}[t]
\begin{center}
\includegraphics[width=\linewidth]{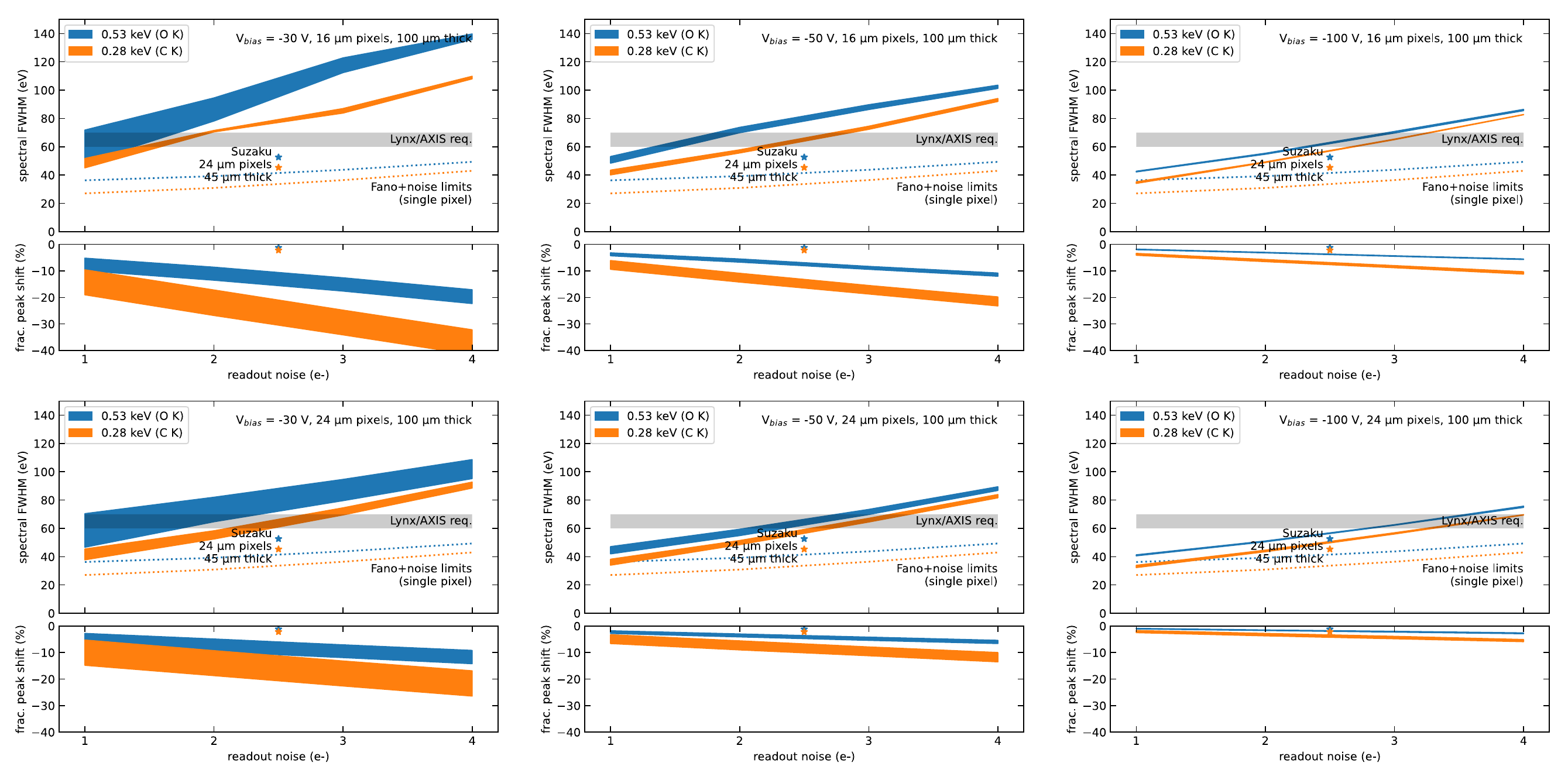}
\end{center}
\caption{Spectral FWHM and peak shift as a function of readout noise for our simulated detector with different pixel size (top: 16 $\mu$m, bottom: 24 $\mu$m) and bias voltage ($V_\mathrm{sub} = -30$ to $-100$ from left to right). Colored bands span a range of high passivation quality, 0--10\% probability of a photoelectron being absorbed if it diffuses to the entrance surface. Theoretical limits are shown by dotted lines, and assume the charge is collected in a single pixel with the specified readout noise. Nominal requirements for a mission like AXIS are shown by the gray band. All notional detectors require noise $<$3 e$^-$ to reach the required FWHM at 0.5 keV; for smaller pixels and lower bias voltage, the noise limits are even more strict. Poorer backside performance will degrade the response further. Performance of the the Suzaku XIS1 CCD is shown as stars at the measured readout noise of 2.5 e$^-$. These CCDs benefit from a small pixel aspect ratio and excellent backside-passivation\cite{Bautzetal2004}, both enabling significantly better charge collection than the thick, small-pixel devices simulated here.}
\label{fig:fwhm_gain}
\end{figure}

Each spectral histogram in Figure \ref{fig:spec} and \ref{fig:spec_mn_mg} was fit with a simple Gaussian to estimate the spectral resolution FWHM and the peak energy shift. These are shown in Figure \ref{fig:fwhm_gain}, with top and bottom panels showing different pixel sizes, and substrate bias voltage increasing from right to left. In effect, the upper left panel is the lower performance detector, and the bottom right panel is the ideal. Colored bands show the range of FWHM and peak shift for a range of ``good'' backside passivation. In all cases, readout noise of 2--3 e$^-$ or lower is necessary to meet the spectral resolution requirements of a mission like AXIS, shown by the gray band. Even at this level, there is little allowance for additional FWHM degradation that may be caused by (for example) poorer backside treatment or charge-transfer inefficiency (CTI) in CCDs\cite{LaMarretal2022b}.

\subsection{Assessing the back surface quality}
\label{sect:passivation}

To maximize QE at X-ray photon energies where the attenuation length in silicon is very short ($<$0.5 $\mu$m below 0.5 keV), X-ray imagers must be thinned, fully depleted, and illuminated on the backside. The thinning process introduces open bonds on the surface that can absorb photoelectrons, and for this and other reasons the surface must be passivated by introducing a thin layer of dopant. The CCID-93 device currently under study uses molecular beam epitaxy (MBE) to introduce this layer\cite{Ryuetal2018}, and its effectiveness at passivating the surface is of great interest as the science goals of future strategic missions demand good soft X-ray QE and spectral resolution.

\begin{figure}[p!]
\begin{center}
\includegraphics[width=\linewidth]{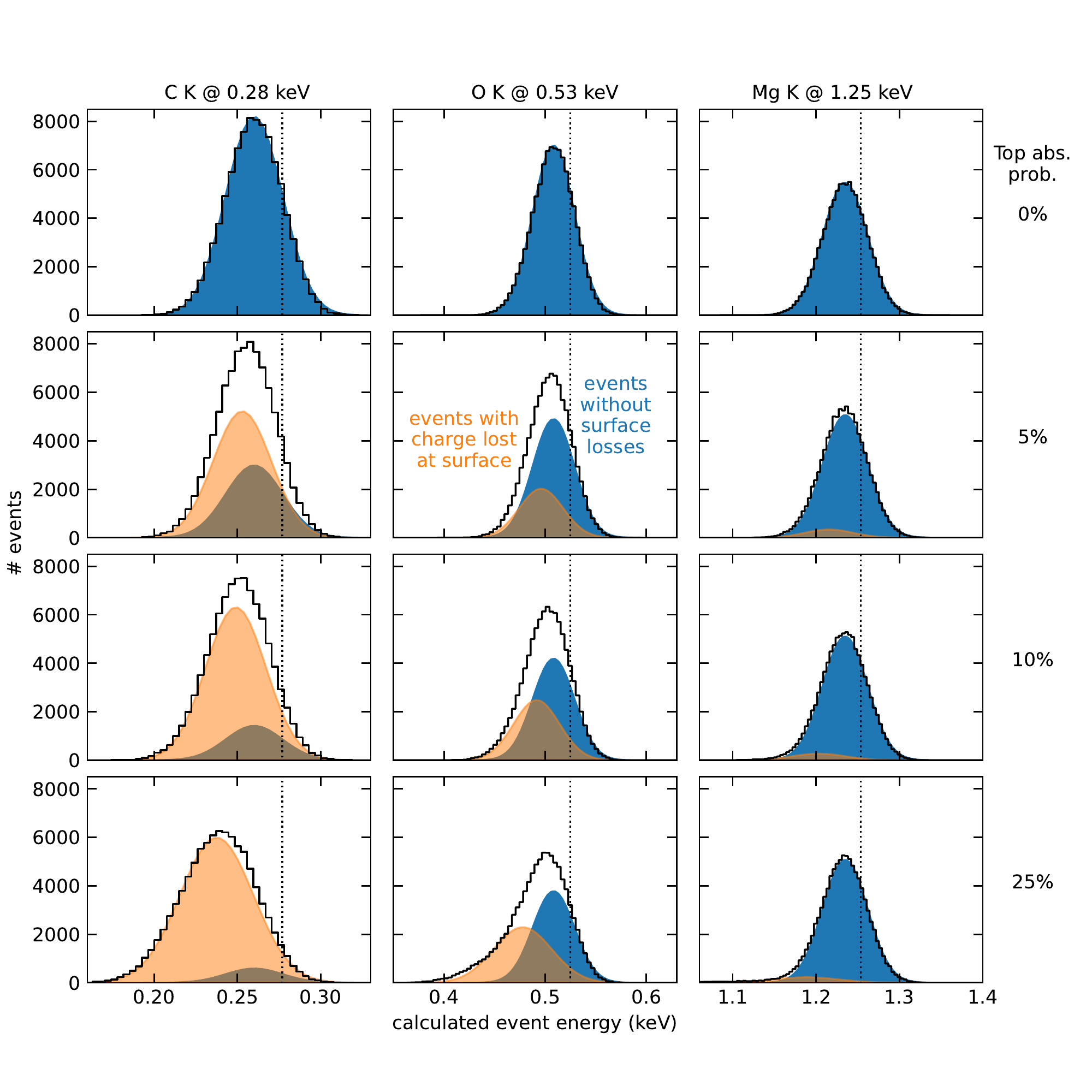}
\end{center}
\caption{Demonstration of a method to assess backside passivation quality by separating it from other response-broadening effects. The black curve in each panel shows the response to monochromatic X-rays from our simulations, with different energies shown in columns and different backside quality shown in rows. Here backside ``quality'' is parameterized by the \textsc{TopAbsorptionProb} probability that a photoelectron encountering the entrance window will be absorbed. We use the simulations with 1 e$^-$ RMS readout noise, 16-$\mu$m pixels, and $V_{\mathrm sub} = -50$ V. The top panels represent perfect passivation, and we fit these histograms with a single Gaussian. The lower panels represent increasingly poor passivation, which we model with two Gaussians as explained in the text. As expected, the fraction of events affected by surface losses (represented by the orange Gaussian) depends strongly on energy.}
\label{fig:passivation}
\end{figure}

As we have shown, a large number of factors contribute to the overall detector response, and these can be difficult to separate from each other in lab data. The simulations we have presented here provide a method for doing just that, since we can easily change the magnitude of these various factors, and since they have been validated for a real-life instrument with exquisite soft response (see Figure \ref{fig:validation}). To demonstrate this, we have used the 1 e$^-$ RMS readout noise, 16-$\mu$m pixel, $V_{\mathrm sub} = -50$ V simulations in an attempt to quantify the fraction of photons that suffer charge loss at the entrance surface. As described in Section \ref{sect:simulations}, the backside quality is captured in \textsc{Poisson CCD} in a parameter called \textsc{TopAbsorptionProb}, the probability that a photoelectron encountering the entrance window will be absorbed and thus lost from the charge packet created by the X-ray. We show in Figure \ref{fig:passivation} monoenergetic spectra for these simulations at three energies (columns) and for four values of the backside quality (rows). Perfect passivation (\textsc{TopAbsorptionProb} = 0\%) is shown at the top, and we fit each of these spectra with a single Gaussian, which is a good fit. For the poorer passivation simulations (\textsc{TopAbsorptionProb} $>$ 0\%), we introduce two Gaussians; the first, shown in blue, represents events that have no surface losses, and in these fits the $\sigma$ and mean are fixed to the perfect passivation Gaussian from the top panel. The second, orange Gaussian represents the ensemble of events that have lost some fraction of charge to the surface, and this is allowed to fit freely. The blue Gaussian amplitude is allowed to vary to account for events that migrate to this ensemble. 

At the lowest energy of 0.28 keV, where the attenuation length is only 0.11 $\mu$m, even a mild decrease in passivation quality results in a majority of photons losing some charge to the surface: 64\% for 5\% absorption probability, and 93\% for 25\%. At 0.53 keV and 0.49 $\mu$m attenuation length, the non-Gaussianity of the response is easier to see, and about half of the photons lose charge to the surface in the 25\% absorption probably case. At 1.25 keV, the attenuation length is $\sim$5 $\mu$m, and while a low-energy tail can be seen, only about 6\% of photons are contained in the orange Gaussian and thus identified as losing charge to the surface. This fraction is the same for all absorption probabilities, and may indicate something physical or a limitation of this technique for photon ensembles which have a range of penetration depths extending well into the silicon bulk. The results for 5.9 keV are not shown, as with a 29-$\mu$m attenuation length there is little surface loss; the vast majority of photons interact away from the surface and have their charge quickly caught up in the bias-induced drift.

We have used the 1 $e^-$ readout noise simulations in this test case because it is easiest to see the effects of backside quality when it dominates the response broadening compared to other factors. In future work, we will expand this method to more demanding cases with higher noise and other factors that degrade resolution, such as CTI. We will also explore a more physical and quantitative method of assessing the backside quality and other surface effects than the simplified \textsc{TopAbsorptionProb} parameter. In particular, we don't expect real events with lost charge to follow a Gaussian distribution, but rather have an extended non-Gaussian or even constant tail to lower energies. The Gaussianity is likely an artifact of this simplified probabilistic charge loss that ignores (for example) initial high-energy photoelectrons completely leaving the silicon. We will also probe the dependence at different energies and as a function of interaction depth of individual photons rather than as an ensemble. 

\section{Summary}
\label{sect:summary}

We have simulated charge diffusion in thick, small-pixel, back-illuminated silicon X-ray imagers to understand the effects on soft X-ray response. The simulations show that, while larger pixels, higher bias voltage, and optimal backside passivation improve performance, reducing the readout noise has a dominant effect in all cases. Since compelling science requirements often compete technically with each other (high spatial resolution, soft X-ray response, hard X-ray response), these results can be used to find the proper balance for a future high-spatial-resolution X-ray instrument.

This variety of response-degrading effects in such detectors complicates assessment of the backside passivation quality, vital for the good soft X-ray response demanded by the science goals of future strategic missions. We have presented a method to qualitatively assess the backside treatment, using our validated simulations to estimate the relative size of the ensembles of photons that lose charge to the surface and those that do not. We will explore this more in future work, with the aim of developing a spectral resolution budget that can be adapted to help formulate future mission requirements.

\acknowledgments 

We gratefully acknowledge support from NASA through the Strategic Astrophysics Technology (SAT) program, grants 80NSSC18K0138 and 80NSSC19K0401 to MIT, and from the Kavli Research Infrastructure Fund of the MIT Kavli Institute for Astrophysics and Space Research.
 
\bibliography{edm} 
\bibliographystyle{spiebib} 

\end{document}